\documentclass[12pt]{article}
\usepackage{amsfonts}
\usepackage{array}
\usepackage{epsfig}
\usepackage{rotating}
\usepackage{amsmath}    
\usepackage{verbatim}   
\usepackage{color}      
\usepackage[dvipsnames]{xcolor}
\usepackage[colorlinks = true,
            linkcolor = blue,
            urlcolor  = blue,
            citecolor = blue,
            anchorcolor = blue]{hyperref}

\usepackage{graphicx}

\usepackage{longtable}
\usepackage{multirow}
\usepackage{multicol}
\usepackage{enumitem}
\usepackage{xspace}

\usepackage{url}

\usepackage{txfonts}
\usepackage[latin1]{inputenc}
\usepackage[T1]{fontenc}

\newcommand{\form}{\textsc{Form}\xspace}

\newcommand{\code}[1]{\texttt{#1}}

\begin{document}


\begin{titlepage}
\thispagestyle{empty}
\noindent
\texttt{\footnotesize\color{Gray}KA-TP-23-2026}
\hfill
September 2026 \\
\noindent
\texttt{\footnotesize\color{Gray}P3H-26-076} \\
\vspace{1.5cm}

\begin{center}
  {\bf \Large
    LLM-Based \form{} Code Generation with \\
    Verification-Driven Fine-Tuning
  }
  \vspace{1.25cm}

 {\large
   Bakar~Chargeishvili\,\footnote{{\tt bakar.chargeishvili@kit.edu}}
 }
 \\
 \vspace{1.25cm}
 {\it
   Karlsruhe Institute of Technology (KIT),
Institut f\"ur Theoretische Physik,\\
Wolfgang-Gaede-Stra{\ss}e 1, 76131 Karlsruhe, Germany\\[1ex]
 }

\vspace{1.4cm}
\large {\bf Abstract}
\vspace{-0.2cm}
\end{center}

\form{} is a domain-specific symbolic manipulation language widely used in
particle physics for processing the very large algebraic expressions arising
from multi-loop Feynman diagram calculations. Despite its central role in
precision theoretical physics, no artificial-intelligence tooling exists, to our
knowledge, for assisting physicists in writing \form{} code. We show that
contemporary large language models (LLMs), including frontier models with
hundreds of billions of parameters, achieve a zero-percent execution pass
rate on our instruction-following and tutorial-style \form{} tasks without
documentation in a single attempt, establishing \form{} as a genuine
zero-shot language for LLMs at the time of writing.
We then present a verification-driven data generation pipeline that uses the
\form{} binary itself as an execution oracle to produce and validate a corpus
of 4{,}633 training examples spanning deterministic computations, open-ended
programs, tutorial code, and knowledge question--answer pairs. Fine-tuning a
compact open-weights model (Qwen3-8B) with quantized low-rank adaptation
(QLoRA) yields a specialist that, evaluated on four complementary benchmarks
(840 tasks, single attempt each), decisively outperforms frontier models
with up to 756B parameters in execution rate and in strict, \form{}-verified
output matching on the larger benchmarks, and remains statistically
indistinguishable from them on the smaller, harder ones. General reasoning
and coding capabilities are preserved within 2.6 percentage points.


\end{titlepage}

\setcounter{footnote}{0}

\newpage





\section{Introduction}
\label{sec:intro}

\form{} is a symbolic manipulation system designed for the efficient
processing of very large algebraic expressions---routinely handling formulae
with millions or even billions of terms that arise in perturbative quantum
field theory calculations~\cite{vermaseren2000form, kuipers2013form,
davies2026form5}. Originally developed by Jos Vermaseren in 1984 as a successor
to Schoonschip~\cite{vermaseren2008form}, \form{} has become an indispensable
tool in particle physics~\cite{manteuffel2025espp}, underpinning calculations of multi-loop Feynman
amplitudes~\cite{larin1996mincer, ruijl2020forcer}, anomalous dimensions,
multiple zeta values~\cite{blumlein2010mzv}, and many other results in
precision QCD and electroweak physics. Its unique architecture---which stores
intermediate expressions on disk rather than in memory---enables it to process
expressions orders of magnitude larger than general-purpose computer algebra
systems such as Mathematica or Maple~\cite{vonhippel2022quanta,
debiasio2026form}.

Despite its importance, \form{} remains challenging to learn and use
effectively. Its syntax differs fundamentally from mainstream programming
languages: pattern matching uses wildcard symbols with set-theoretic
conditions, the \code{.sort} statement controls expression buffering, and
tensor operations require explicit index declarations. The learning curve is
long, and there is no ecosystem of AI-assisted coding tools comparable to what
exists for Python, JavaScript, or even specialised languages like SQL.

The emergence of large language models (LLMs) trained on code---exemplified by
Codex~\cite{chen2021codex} and its successors---has transformed software
development for mainstream languages. However, \form{} is a \emph{zero-resource
language} in the LLM training-data sense: it appears negligibly or not at all
in the internet-scale corpora on which modern LLMs are trained. Throughout,
\emph{zero-resource} refers to this training-data property; we reserve
\emph{zero-shot} for the absence of in-context documentation or examples
(Section~\ref{sec:baselines}). This raises a
fundamental question: \emph{can a small language model be specialised to
generate correct \form{} code, and how does it compare to frontier-scale
models?}

In this work, we address this question through three contributions:

\begin{enumerate}[leftmargin=*,itemsep=2pt]
  \item \textbf{A verification-driven data generation pipeline} that exploits
  the \form{} binary as an execution oracle. The pipeline generates candidate
  programs from natural-language instructions using a frontier LLM, then
  verifies each program through a four-stage process: deterministic-word
  checking, syntax validation via the \form{} binary, non-trivial output
  confirmation, and semantic consistency auditing by an independent LLM. This
  yields 4{,}633 verified training examples with zero train/test overlap.

  \item \textbf{A parameter-efficient fine-tuning recipe} based on QLoRA
  \cite{dettmers2023qlora, hu2022lora} applied to the 8-billion-parameter
  Qwen3-8B model~\cite{yang2025qwen3}. The resulting model achieves 97.7\%
  exact-output-match on a 664-task deterministic benchmark and 83.0\%
  execution rate on a 100-task open-ended benchmark, outperforming every
  frontier model studied---including GLM-5.2 and GLM-5.3 (756B
  parameters~\cite{zeng2026glm5, glm5team2026glm53}) and DeepSeek-V4 Flash (304B
  parameters~\cite{xu2026deepseekv4})---with non-overlapping 95\% confidence
  intervals on both benchmarks. The advantage persists under a strict,
  \form{}-verified output-match metric (18.0\% vs.\ 5.6\% for the best
  frontier model on 89 of the 100 tasks whose output is uniquely
  determined).

  \item \textbf{A comprehensive evaluation framework} comprising four
  benchmarks, seven baseline models, execution-based metrics, anchored LLM
  judging with a \form{}-specific factsheet, 95\% bootstrap confidence
  intervals, and a strict \form{}-verified output-match metric that separates
  genuinely solved problems from merely runnable programs. We also demonstrate that general reasoning capability is
  preserved ($-2.6$\,pp on MMLU~\cite{hendrycks2021mmlu}, $-1.5$\,pp on
  GSM8K~\cite{cobbe2021gsm8k}), confirming that QLoRA fine-tuning does not
  cause catastrophic forgetting.
\end{enumerate}

The remainder of this paper is organised as follows. In Section~\ref{sec:background}
we review the \form{} symbolic manipulation system, language models for code
generation, and parameter-efficient fine-tuning. Section~\ref{sec:method}
describes our verification-driven data generation pipeline and the fine-tuning
configuration. In Section~\ref{sec:evaluation} we introduce the benchmark
suite, baseline models, and evaluation protocol. Section~\ref{sec:results}
presents our findings, including the zero-shot result, main benchmark results,
ablation studies, and capability preservation. We discuss why a small
fine-tuned model outperforms frontier models, the limitations of the present
work, and future directions in Section~\ref{sec:discussion}, and conclude in
Section~\ref{sec:conclusion}.


\section{Background}
\label{sec:background}

\subsection{The \form{} Symbolic Manipulation System}
\label{sec:form}

\form{} is a domain-specific language (DSL) for symbolic algebra, designed
specifically for the kind of expression manipulation that arises in
perturbative quantum field theory. Unlike general-purpose computer algebra
systems, \form{} is optimised for \emph{throughput} rather than interactivity:
it processes expressions in a batch-sequential pipeline, writing intermediate
results to disk and sorting them between processing stages via the
\code{.sort} command~\cite{kuipers2013form}.

A typical \form{} program declares indices, vectors, and tensors;
defines a local expression; applies trace and contraction operations; and
prints the result. Figure~\ref{fig:form-example} shows a simple example that
computes the Dirac trace of two gamma matrices contracted with external
vectors.

\begin{figure}[ht]
\begin{verbatim}
Indices alpha, beta;
Vectors k, l;
Local E = g_(1, alpha, beta) * k(alpha) * l(beta);
Trace4, 1;
contract;
Print;
.end
\end{verbatim}
\caption{A simple \form{} program: declare indices and vectors, define an
expression with Dirac gamma matrices $g_(1,\alpha,\beta)$, take the trace
(\code{Trace4,\,1}), contract Lorentz indices, and print. The output is
$E = 4\,k\cdot l$, reflecting the trace identity
$\mathrm{Tr}[\gamma^\alpha \gamma^\beta] = 4\,g^{\alpha\beta}$.}
\label{fig:form-example}
\end{figure}

Key features that make \form{} challenging for LLMs include:
\begin{itemize}[leftmargin=*,itemsep=1pt]
  \item \textbf{Wildcard pattern matching}: Wildcard variables must be
  declared as \code{Symbols} and use the \code{?} syntax (e.g.,
  \code{id f(x?,y?) = g(x,y)}), with optional set conditions
  (e.g., \code{\{>2,<5\}}).
  \item \textbf{Tensor operations}: The metric tensor \code{d\_}, the
  Levi-Civita symbol \code{e\_}, and Dirac gamma matrices \code{g\_} require
  explicit index declarations and dimension settings. The trace operation
  \code{Trace4} acts on gamma-matrix chains.
  \item \textbf{Preprocessor loops}: The \code{\#do}/\code{\#enddo} and
  \code{\#if} constructs enable metaprogramming, but loop variables in
  conditions require backtick syntax (e.g., \code{\#if `i' > 2}).
  \item \textbf{Expression lifecycle}: Within a single module, \code{Local F = F + \ldots}
  \emph{redefines} rather than accumulates; iterative accumulation inside
  \code{\#do} loops requires an intervening \code{.sort} between loop
  iterations.
  \item \textbf{No built-in calculus}: \form{} has no \code{diff\_()},
  \code{Int()}, or \code{det()} functions. Differentiation, integration, and
  determinant computation must be implemented manually via pattern matching.
\end{itemize}

\form{} has been used in landmark physics calculations including the
three-loop QCD anomalous dimensions~\cite{larin1996mincer}, four-loop
massless propagator integrals via the Forcer package~\cite{ruijl2020forcer},
and the systematic computation of multiple zeta values~\cite{blumlein2010mzv}.
The 2022 \emph{Quanta Magazine} article ``Crucial Computer Program for
Particle Physics at Risk of Obsolescence''~\cite{vonhippel2022quanta}
highlighted both \form{}'s irreplaceability and its vulnerability as a
single-maintainer project; a recent \emph{CERN Courier} interview with
\form{}'s past, present and future developers discusses the same
sustainability concerns~\cite{debiasio2026form}. Our work aims to lower
the barrier to \form{} proficiency by providing an AI coding assistant,
thereby broadening the user
base and partially mitigating the knowledge-transfer problem.

\subsection{Language Models for Code Generation}
\label{sec:llm-code}

The application of LLMs to code generation has been extensively studied for
mainstream languages. The HumanEval benchmark~\cite{chen2021codex}
introduced execution-based evaluation with the pass@k metric, now standard
across the field. For low-resource and domain-specific languages, the
challenge is acute: Cassano et al.~\cite{cassano2024multiplt} proposed
MultiPL-T, a pipeline for translating high-resource training data to
low-resource languages, demonstrating that fine-tuning on synthetic data can
significantly improve performance on languages with minimal public code
repositories. Mora et al.~\cite{mora2024synthetic} introduced ``synthetic programming elicitation'' for very low-resource languages, using formal grammar
constraints to generate and validate training data.

Our approach differs from these works in two key respects. First, we benefit
from having a \emph{deterministic execution oracle} (the \form{} binary) that
provides ground-truth verification of both syntax and output---a stronger
signal than the unit-test-based verification used in
MultiPL-T~\cite{cassano2024multiplt} or the grammar-based validation
in~\cite{mora2024synthetic}. Second, \form{} is not merely a low-resource
language but a \emph{zero-shot language}: as we demonstrate in
Section~\ref{sec:results}, all tested LLMs (up to 756B parameters) produce
zero syntactically-valid \form{} programs without documentation context on
Instruct-100, and the same holds on Tutorial-44 for the base model and the
frontier models evaluated there without documentation
(Section~\ref{sec:zero-shot} details the no-documentation coverage).

Our generation-and-verification pipeline is conceptually related to the
\emph{Ralph Wiggum Loop} (RWL)~\cite{xu2026rwl}, an agentic design pattern in
which an LLM repeatedly generates candidate solutions, an external verifier
evaluates them, and the LLM reflects on the feedback to produce improved
candidates. In the RWL framework---named after the \textit{Simpsons} character
for its persistent trial-and-error behaviour---the agent alternates between
generation and external validation until the solution passes all checks or a
maximum iteration count is reached. Our pipeline instantiates this pattern
with the \form{} binary as the external verifier: candidate programs are
generated by a frontier LLM, validated through execution, and the error
output is fed back for repair attempts. The key difference is that our
pipeline collects only the \emph{successfully verified} outputs as training
data, rather than using the iterative refinement to improve the generator
itself.

\subsection{Parameter-Efficient Fine-Tuning}
\label{sec:peft}

Full supervised fine-tuning (SFT) of an 8B-parameter model must keep the
full parameter set, gradients, and optimizer states in memory---on the order
of 70--80\,GB with AdamW---which is feasible on
modern hardware but risks \emph{catastrophic forgetting} of pre-trained
capabilities. QLoRA~\cite{dettmers2023qlora} addresses both concerns by
quantising the base model to 4-bit precision and training only low-rank
adapter matrices (LoRA~\cite{hu2022lora}) on top of the frozen weights.
This reduces the number of trainable parameters by $\sim$99\% while
preserving performance comparable to full fine-tuning.


\section{Method}
\label{sec:method}

\subsection{Verification-Driven Data Generation}
\label{sec:data-pipeline}

\begin{figure}[t]
  \centering
  \includegraphics[width=\textwidth]{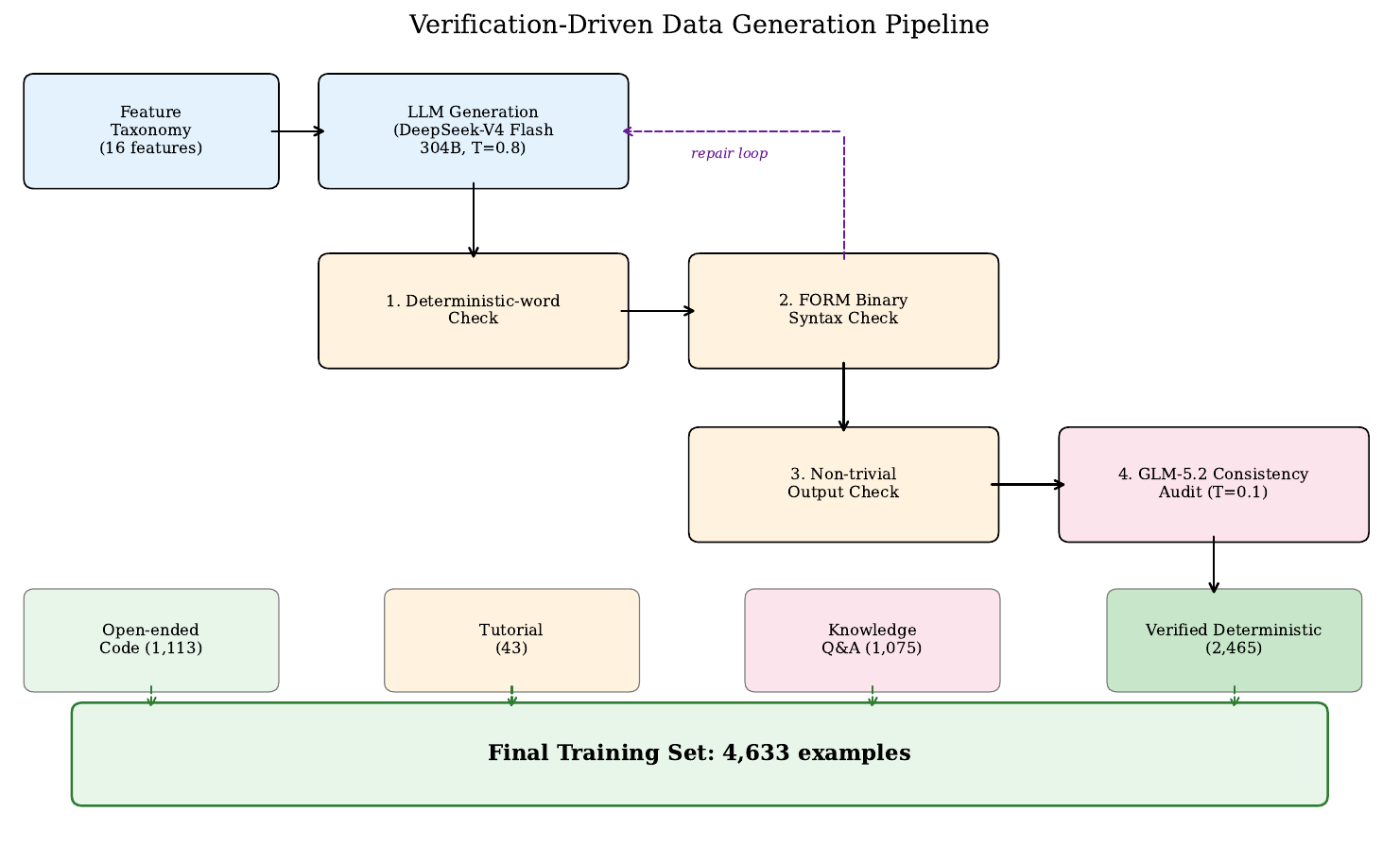}
  \caption{Verification-driven data generation pipeline. Candidate programs
  are generated from natural-language instructions by a frontier LLM
  (DeepSeek-V4 Flash, temperature\,0.8), then pass through a four-stage
  verification process before inclusion in the training set.}
  \label{fig:pipeline}
\end{figure}

Our data generation pipeline is built on the principle that every training
example must be verified by executing the generated \form{} program through
the actual \form{} binary. This ensures that the model learns only from
syntactically valid, semantically consistent examples. The pipeline consists
of four stages, illustrated in Figure~\ref{fig:pipeline}.

\subsubsection*{Feature Taxonomy}
We define a taxonomy of 16 \form{} features, each with three complexity levels
(basic, intermediate, advanced), yielding 48 feature-complexity cells. The
features include: symbol and expression arithmetic, index and tensor
operations, gamma-matrix traces, Levi-Civita contractions, wildcard pattern
matching, preprocessor loops (\code{\#do}/\code{\#if}), expression
manipulation (\code{repeat}/\code{while}), the \code{count} function, the
\code{discard}/\code{keep} mechanism, nested function substitution, and
polynomial operations.

\subsubsection*{Generation}
For each feature--complexity cell, we generate candidate programs by prompting
DeepSeek-V4 Flash (304B parameters~\cite{xu2026deepseekv4}) at temperature 0.8
with instructions in plain English. The instructions specify the exact
symbols, expressions, and operations to use, ensuring deterministic outputs.
This produces $\sim$10{,}000 candidate programs across all cells.

\subsubsection*{Four-Stage Verification}
Each candidate program passes through the following verification chain:

\begin{enumerate}[leftmargin=*,itemsep=2pt]
  \item \textbf{Deterministic-word check}: The instruction is scanned for
  non-deterministic words (``think'', ``appropriate'', ``suitable'') that
  could lead to variable outputs. Candidates containing such words are
  rejected.

  \item \textbf{\form{} binary syntax validation}: The program is executed
  through the \form{} binary (FORM\,5.0, April\,2025). Programs that fail to
  compile or run are rejected. Importantly, \form{} writes error messages to
  standard output (not standard error) with exit code\,1, which we capture for
  the repair loop.

  \item \textbf{Non-trivial output check}: Programs that produce empty or
  trivially simple output (e.g., \code{0} or \code{1}) are rejected to ensure
  training signal richness.

  \item \textbf{LLM consistency audit}: An independent LLM (GLM-5.2, 756B
  parameters~\cite{zeng2026glm5}, temperature\,0.1; low temperature since
  auditing is deterministic classification, not creative
  generation---cf.\ the LLM judge of Section~\ref{sec:metrics}) evaluates
  whether the generated program is semantically consistent with the
  instruction. This
  catches cases where the program runs but does not implement the requested
  computation. The auditor is given a \emph{\form{} factsheet}---a curated list
  of real \form{} built-in functions and common pitfalls---to prevent
  hallucinated capabilities from inflating audit scores.
\end{enumerate}

After verification, duplicate programs are removed via exact-match and
normalised-text deduplication. The final deterministic training set contains
2{,}465 verified examples.

\subsubsection*{Multi-Source Data Composition}

Beyond deterministic tasks, we supplement the training set with three
additional data categories:

\begin{itemize}[leftmargin=*,itemsep=2pt]
  \item \textbf{Open-ended code generation} (1{,}050 training examples): Instructions
  that specify a goal but leave implementation details open. These are
  generated from seed programs in the \form{} reference manual and the
  ``\form{} for Pedestrians'' tutorial~\cite{heck2000pedestrians}, then
  GLM-5.2-audited for instruction--code consistency (63 of the 1{,}113
  generated examples are held out for validation).

  \item \textbf{Tutorial programs} (43 examples): Verbatim programs extracted
  from the ``\form{} for Pedestrians'' book~\cite{heck2000pedestrians}, paired
  with natural-language instructions. These provide canonical idioms that
  reflect how \form{} is actually used in practice. The 43 training programs
  are exact-match-disjoint from the 44 Tutorial-44 test tasks.

  \item \textbf{Knowledge question--answer pairs} (1{,}075 examples):
  Conceptual questions about \form{} syntax and semantics (e.g., ``What does
  the \code{.sort} statement do?''), generated by an LLM and audited by an
  independent LLM judge.

  \item \textbf{Documentation-in-context slice} (355 rows): A variant subset
  of the training data---counted within the categories above, not in addition
  to them---where the first 6{,}000 characters of a \form{} syntax guide
  are prepended to the instruction (applied to 10\% of the code rows),
  teaching the model to use documentation at
  inference time.
\end{itemize}

The final training set therefore comprises 2{,}465 deterministic +
1{,}050 open-ended + 43 tutorial + 1{,}075 question--answer examples
$= 4{,}633$ examples with a 314-example
validation split. Test sets are generated \emph{before} training data to
prevent data leakage; we verify zero exact-match contamination between train
and test instructions.

\subsection{Fine-Tuning Configuration}
\label{sec:training}

We fine-tune Qwen3-8B~\cite{yang2025qwen3}, a dense model with 8 billion
parameters, using QLoRA with the configuration in Table~\ref{tab:hyperparams}.

\begin{table}[t]
\centering
\caption{QLoRA fine-tuning hyperparameters.}
\label{tab:hyperparams}
\renewcommand{\arraystretch}{1.45}
\begin{tabular}{ll}
\hline\hline
Parameter & Value \\
\hline
Base model & Qwen3-8B (4-bit NF4 quantisation) \\
LoRA rank ($r$) & 16 \\
LoRA alpha ($\alpha$) & 32 \\
LoRA dropout & 0.05 \\
Target modules & All linear layers \\
Learning rate & $2 \times 10^{-4}$ \\
LR schedule & Cosine, warmup 5\% \\
Epochs & 5 \\
Effective batch size & 32 (8 per device $\times$ 4 grad.\ accum.) \\
Max sequence length & 2{,}048 tokens \\
Optimizer & AdamW (paged) \\
Framework & Unsloth~\cite{unsloth2024} + TRL~\cite{trl2024} \\
Training time & $\sim$35 minutes (1$\times$ A100-40GB) \\
\hline\hline
\end{tabular}
\end{table}

We train two model variants to study the effect of data composition:
\begin{itemize}
  \item \textbf{FORM-8B (det-only)}, hereafter \textbf{v3b}: 2{,}465
  deterministic examples only. This model serves as a \emph{negative control}:
  it demonstrates what happens when the training data is verified but
  compositionally narrow (deterministic tasks with fully specified
  instructions and known outputs).
  \item \textbf{FORM-8B (full mix)}, hereafter \textbf{v3c}: 4{,}633 examples
  including deterministic tasks, open-ended code generation, tutorial
  programs, knowledge Q\&A pairs, and the documentation-in-context slice.
  This is our primary model. The label ``v3c'' reflects the third iteration
  of our data pipeline, incorporating audited multi-source data.
\end{itemize}
The rationale for training two variants is to isolate the effect of data
\emph{composition} from data \emph{verification}: both v3b and v3c use verified
data, but v3b contains only deterministic tasks while v3c includes diverse
open-ended, tutorial, and knowledge components. As we show in
Section~\ref{sec:ablation}, this difference in composition is the single most
important factor for open-ended capability.

\begin{figure}[t]
  \centering
  \includegraphics[width=\textwidth]{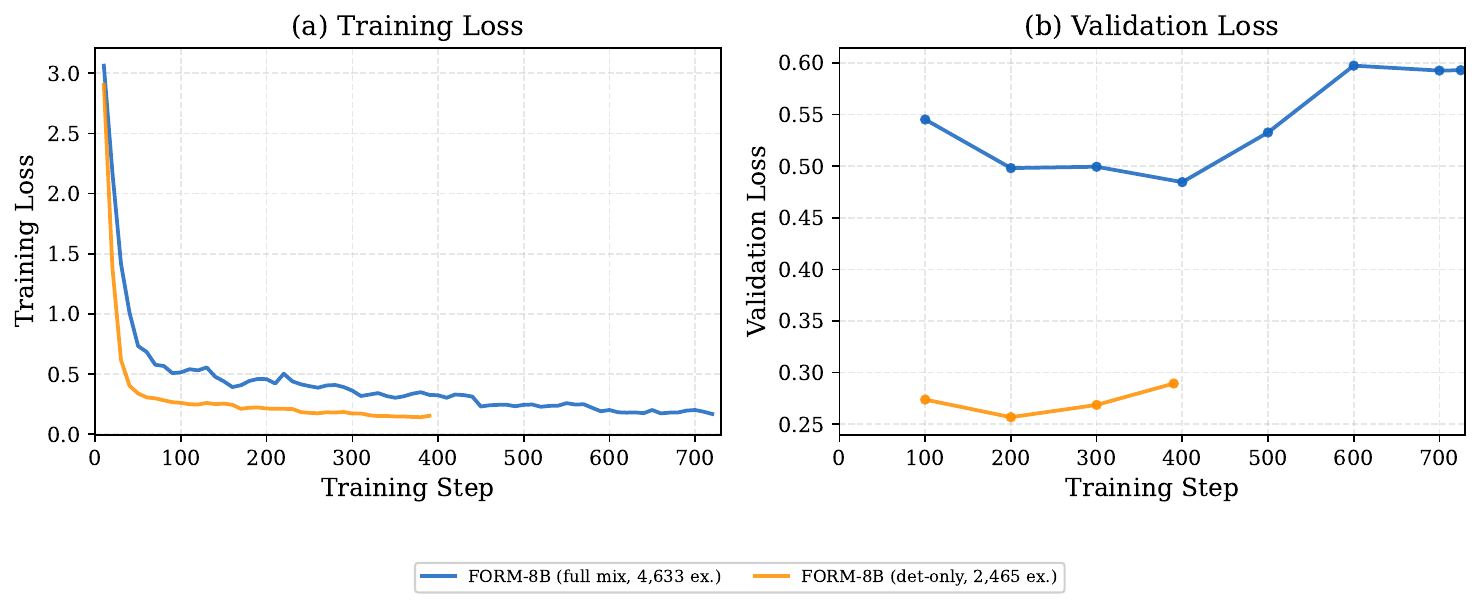}
  \caption{Training (left) and validation (right) loss curves for the two
  model variants: v3b (det-only, 2{,}465 examples) and v3c (full mix, 4{,}633
  examples). The v3c model converges to a higher validation loss due to
  task diversity but achieves substantially better performance on open-ended
  benchmarks (see Section~\ref{sec:results}).}
  \label{fig:training}
\end{figure}

Training loss curves are shown in Figure~\ref{fig:training}. Both variants
converge smoothly. The det-only model achieves lower validation loss (0.289)
due to the narrower output distribution, while the full-mix model stabilises
at $\sim$0.59, reflecting the broader task variety.

We disable Qwen3's hybrid thinking mode during both training and inference,
such that the model directly emits \form{} code without producing an
intermediate reasoning trace. This is a deliberate design choice: the training
data contains only code (no chain-of-thought), and we prioritise fast,
deterministic code generation over explainability. The model has nevertheless
\emph{internalised} \form{} syntax conventions---for example, it consistently
uses \code{Trace4} for Dirac traces, respects the redefinition semantics of
\code{Local}, and applies \code{.sort} before cross-expression references---
suggesting that these patterns are learned as tacit knowledge rather than
explicit reasoning steps.

\section{Evaluation}
\label{sec:evaluation}

\subsection{Benchmark Suite}
\label{sec:benchmarks}

We evaluate on four benchmarks, designed to cover different aspects of
\form{} code generation:

\begin{enumerate}[leftmargin=*,itemsep=2pt]
  \item \textbf{B1: 664-Deterministic} (664 tasks). Fully-specified
  computations with known expected outputs. Metric: exact output match after
  running the generated program through the \form{} binary. This is the
  primary benchmark; its large size (664) yields tight confidence intervals.

  \item \textbf{B2: Instruct-100} (100 tasks). Open-ended instructions that
  specify a computation goal without prescribing the exact implementation.
  Metric: syntax pass (the generated program compiles and runs without error
  in the \form{} binary), plus a strict \form{}-verified output-match rate
  on the 89 output-determined tasks (eleven references print no result
  expression; Section~\ref{sec:true-solve}); the
  latter is a conservative lower bound because many instructions
  underspecify the target computation.

  \item \textbf{B3: Tutorial-44} (44 tasks). Programs extracted from the
  ``\form{} for Pedestrians'' tutorial~\cite{heck2000pedestrians}, with
  instructions re-written in plain English. This benchmark is
  \emph{completely independent} of the training pipeline and tests
  generalisation to canonical \form{} idioms. Metric: syntax pass, plus the
  same strict \form{}-verified output-match rate on the 41 adjudicable
  tasks (three references print no result expression).

  \item \textbf{B4: Open-32} (32 tasks). Complex, multi-step \form{}
  tasks (e.g., ``Compute the SU(3) Casimir operator and verify its
  eigenvalue''). Metric: PASS/PARTIAL/FAIL verdict from an anchored LLM judge
  (GLM-5.2, temperature\,0.1) that scores each task on 2--6 verifiable
  requirements, with a \form{}-specific factsheet in the system prompt to
  prevent hallucinated capabilities. The judge ceiling is verified at 100\%
  (32/32 reference solutions judged PASS). In addition, the 22
  output-determined tasks are graded by strict \form{}-verified output
  matching (Section~\ref{sec:true-solve}).
\end{enumerate}

All benchmarks use a \emph{zero-shot, single-attempt} protocol: the model
receives the instruction (with no few-shot examples) and produces exactly one
generation, which is then evaluated. This is the strictest fair-evaluation
protocol. The single-attempt design is a deliberate choice of eval target:
real coding practice is of course iterative, but agentic loops measure tool
orchestration as much as language competence, are strongly sensitive to
system prompts and tooling, and are run-to-run stochastic. A single attempt
at temperature 0 measures the one competence every such loop still
requires---first-shot generation from a problem statement alone---in a way
that is deterministic and exactly reproducible. We demonstrate the
specialist inside a realistic agentic workflow separately
(Section~\ref{sec:agentic}); a fair cross-model agentic benchmark is beyond
the scope of this work.

\subsection{Baseline Models}
\label{sec:baselines}

We compare against seven baseline models, representing a range of scales and
architectures (Table~\ref{tab:baselines}). ``Docs'' indicates that the first
6{,}000 characters of a \form{} syntax guide are prepended to the
instruction. Without documentation, every frontier model we evaluated
produces zero valid programs on Instruct-100, and on Tutorial-44 the base model,
DeepSeek-V4 Flash, GPT-OSS-120B and Gemma 4 likewise produce zero
(Section~\ref{sec:zero-shot} gives the Open-32 exception).

\begin{table}[t]
\centering
\caption{Baseline models. ``Docs'' indicates that the first 6{,}000 characters
of a \form{} syntax guide are prepended to the instruction. Without
documentation, all evaluated frontier models score 0\% on Instruct-100, and
the base model, DeepSeek-V4 Flash, GPT-OSS-120B and Gemma 4 likewise score
0\% on Tutorial-44 (Open-32, whose rubric judge credits partial solutions,
is the one exception; see Section~\ref{sec:zero-shot}). GLM-5.3 shares
the GLM-5.2 base model (756B total, 40B active); its gains over GLM-5.2 are
post-training only~\cite{glm5team2026glm53}.}
\label{tab:baselines}
\renewcommand{\arraystretch}{1.3}
\begin{tabular}{lrrll}
\hline\hline
Model & Params & Active & Type & Context \\
\hline
Qwen3-8B (base)~\cite{yang2025qwen3} & 8B & 8B & Dense & Zero-shot \\
FORM-8B (ours) & 8B & 8B & Dense + QLoRA & Zero-shot \\
Qwen3.8-27B~\cite{qwen2026qwen38} & 27B & 27B & Dense & +Docs \\
Gemma 4~\cite{deepmind2026gemma4} & 33B & 33B & Dense & +Docs \\
GPT-OSS-120B~\cite{agarwal2025gptoss} & 117B & 5.1B & MoE & +Docs \\
DeepSeek-V4 Flash~\cite{xu2026deepseekv4} & 304B & 13B & MoE & +Docs \\
GLM-5.2~\cite{zeng2026glm5} & 756B & 40B & MoE & +Docs \\
GLM-5.3~\cite{glm5team2026glm53} & 756B & 40B & MoE & +Docs \\
\hline\hline
\end{tabular}
\end{table}

All frontier models are evaluated \emph{with} documentation (the first
6{,}000 characters of a curated \form{} syntax guide prepended to the
instruction), because without it they score 0\%---a finding we report in
Section~\ref{sec:results}. This provides frontier models with a significant
advantage: our fine-tuned model is evaluated zero-shot (no documentation),
while frontier models receive documentation that covers most of the syntax
needed for the benchmarks.

\subsection{Metrics and Statistical Analysis}
\label{sec:metrics}

For the deterministic benchmark (B1), we use exact output match: the
generated program is executed through the \form{} binary, and its stdout is
compared against the expected output.

For syntax-pass benchmarks (B2, B3), we check whether the generated program
compiles and runs without error in the \form{} binary.

For the open-ended benchmark (B4), we use an anchored LLM judge (GLM-5.2,
temperature\,0.1) that scores each task on 2--6 \emph{verifiable criteria}
(e.g., ``correct Levi-Civita sign'', ``proper Trace4 usage''). The judge is
sampled at a deliberately low temperature rather than the vendor default
($\approx$0.6, tuned for conversational diversity): rubric scoring is a
deterministic classification task, where sampling noise only adds verdict
variance. The only other parameters overridden are a generation-length cap
and disabling the model's thinking mode; all remaining sampling parameters
use the server defaults. The judge's accuracy does not rely on creative
diversity---its reference ceiling is 32/32 (all reference solutions judged
PASS), and its verdicts are largely confirmed by the \form{}-verified ground
truth on the output-determined tasks (11 of 11 rubric passes for our model;
Section~\ref{sec:true-solve}). The score is
the fraction of criteria met (0--1), with the verdict derived as follows:
\begin{itemize}[leftmargin=*,itemsep=1pt]
  \item PASS\,$\Leftrightarrow$\,score\,=1.0,
  \item PARTIAL\,$\Leftrightarrow$\,0\,<\,score\,<\,1,
  \item FAIL\,$\Leftrightarrow$\,score\,=0 or no valid run.
\end{itemize}
The judge's system prompt includes a
\emph{\form{} factsheet} listing real built-in functions and common pitfalls
(e.g., \form{} has no \code{diff\_()} or \code{det()} function; \code{e\_}
contraction yields $+24$ in FORM's convention), preventing the judge from
crediting hallucinated capabilities. The exact factsheet text used for the
published rubric verdicts is archived in the benchmark release
(\code{judge\_factsheet\_v1.txt}); the accumulation pitfall it encodes is
discussed, in corrected form, in Section~\ref{sec:background}.

We report 95\% bootstrap confidence intervals (10{,}000 resamples) for all
benchmarks. Significance between models is assessed by checking for
non-overlapping confidence intervals---a conservative criterion. For
zero-count entries the bootstrap percentile interval is degenerate at zero;
we substitute the one-sided 95\% Clopper--Pearson upper bound.
As a complementary and more powerful criterion we additionally compare
models pairwise with the exact version of McNemar's
test~\cite{mcnemar1947note}. Because all models attempt the same tasks, the
comparison reduces to the \emph{discordant} pairs---tasks solved by one
model but not the other---whose split is tested against a fair-coin null
hypothesis with an exact two-sided binomial test. Pairing controls for task
difficulty, which CI overlap does not; its cost is that discordant pairs
alone carry the signal, so tiny benchmarks yield little power. We use it to
check which of the differences suggested by Table~\ref{tab:main-results} are
statistically real.

In addition to these execution-based metrics, we report a \emph{strict
output-match} rate for B2--B4: the generated program is re-executed and its
result expressions must reproduce the reference output, with equality
verified by \form{} itself (Section~\ref{sec:true-solve} gives the full
procedure). This separates programs that \emph{run} from programs that
\emph{solve} the stated problem, and is the correctness metric emphasised
in Table~\ref{tab:main-results}.


\section{Results}
\label{sec:results}

\subsection{\form{} is a True Zero-Shot Language}
\label{sec:zero-shot}

Our first finding is that \form{} is a \emph{zero-shot language} for all
the LLMs evaluated without documentation: every model---from the 8B base
model to the 756B GLM-5.2---produces \emph{zero} syntactically-valid \form{}
programs without documentation on Instruct-100, and likewise on Tutorial-44
for every model evaluated there (base, DeepSeek-V4 Flash, GPT-OSS-120B and
Gemma 4). The rubric-judged Open-32
benchmark is the one exception: without documentation, the largest frontier
models still complete a non-trivial fraction of its tasks (GLM-5.3 46.9\%,
GLM-5.2 40.6\%, DeepSeek-V4 Flash 37.5\%; GPT-OSS-120B 12.5\% and Gemma 4
6.2\% remain near zero). Strict \form{}-verified adjudication
(Section~\ref{sec:true-solve}) confirms that this competence is genuine
rather than an artefact of the rubric judge: without documentation,
GLM-5.3, GLM-5.2 and DeepSeek-V4 Flash genuinely solve 7, 8 and 9 of the 22
output-determined Open-32 tasks, respectively. The contrast with the
Instruct-100 and Tutorial-44 benchmarks is explained by task shape: Open-32
problems are fully self-contained and closer to generic symbolic
programming, whereas B2/B3 hinge on \form{}-specific idioms. Taken
together, the near-zero no-docs performance indicates that \form{}
appears negligibly in the training corpora of contemporary LLMs, placing it
in the category of zero-resource languages as defined by Cassano
et al.~\cite{cassano2024multiplt}. All frontier-model results reported
subsequently are \emph{with} documentation context, providing them with a
substantial advantage over our zero-shot fine-tuned model.

\subsection{Main Results}
\label{sec:main-results}

\begin{figure}[t]
  \centering
  \includegraphics[width=\textwidth]{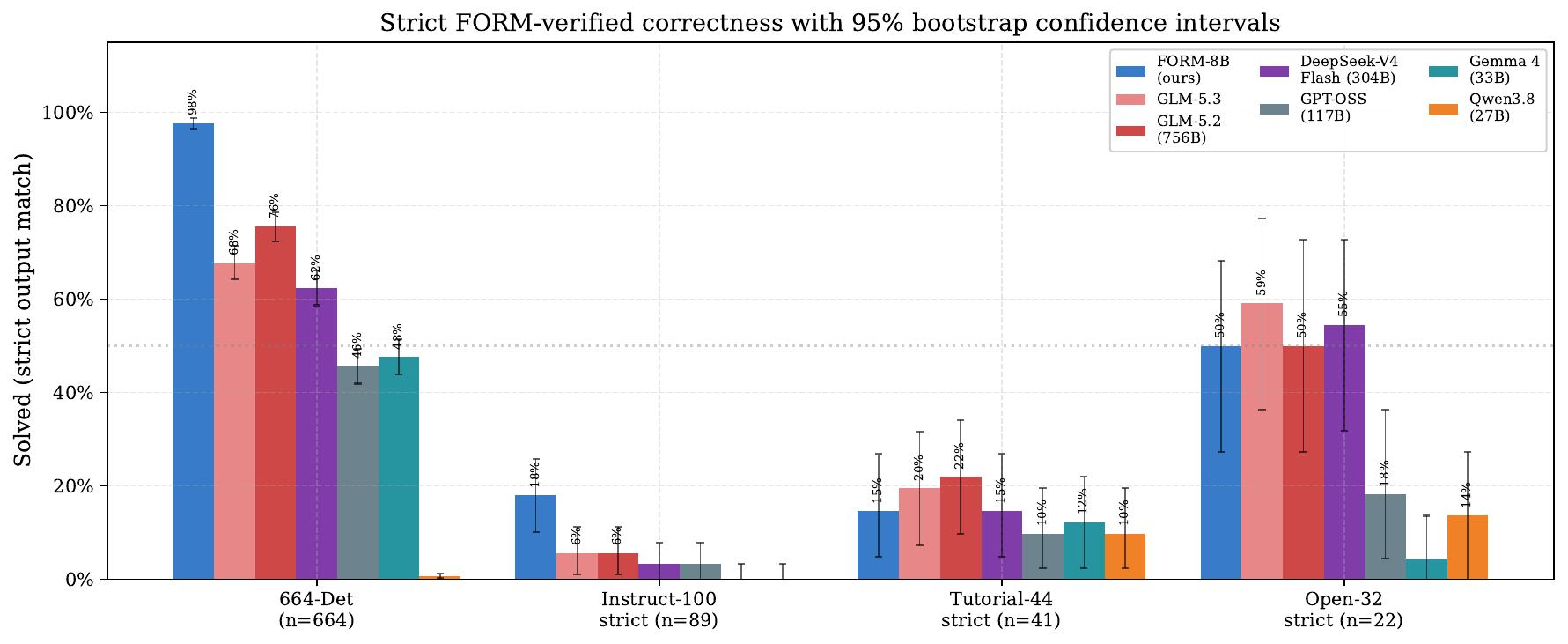}
  \caption{Strict \form{}-verified correctness rates (Table~\ref{tab:main-results},
  ``Strict'' columns; B1 is output match by construction) with 95\% bootstrap
  confidence intervals. Our fine-tuned 8B model (blue) dominates on B1 and
  on the output-determined B2 tasks; on B3 and B4 the small adjudicable
  samples (41, 22)
  produce wide CIs with overlap.}
  \label{fig:main}
\end{figure}

Table~\ref{tab:main-results} presents the results across all four
benchmarks. For B2--B4 each benchmark occupies two columns: the execution
rate (syntax pass; rubric-judged for B4's free-choice tasks) and the
\emph{strict} \form{}-verified correctness rate described in
Section~\ref{sec:true-solve}, which is the metric we emphasise.
Figure~\ref{fig:main} visualises the strict rates with confidence
intervals.

\begin{table}[tp]
\centering
\caption{Main results across the four benchmarks, reported as
execution rate (``Exec.'': syntax pass for B2/B3, rubric score for B4) and
\form{}-verified strict output-match rate (``Strict''). B1 is strict output
match by construction (n=664). The B2 strict rate is computed on the 89
output-determined tasks and is a conservative lower bound
(Section~\ref{sec:true-solve}); the B4 strict rate on the 22
output-determined tasks (for GPT-OSS-120B and Gemma~4 it is based on the
no-docs runs); the B3 strict rate on the 41 adjudicable tasks (three
references print no result expression and cannot be verified by
output matching). Frontier models are evaluated \emph{with} documentation; our
model is \emph{zero-shot}. Bold indicates the best score per column. 95\%
bootstrap CIs for the correctness metrics (B1 and all Strict columns).
All entries are single-attempt generations at temperature 0.}
\label{tab:main-results}
\small
\renewcommand{\arraystretch}{1.45}
\setlength{\tabcolsep}{1.8pt}
\begin{tabular}{lccccccc}
\hline\hline
Model & B1 & \multicolumn{2}{c}{B2: Instruct-100} & \multicolumn{2}{c}{B3: Tutorial-44} & \multicolumn{2}{c}{B4: Open-32} \\
 & (n=664) & Exec. & Strict & Exec. & Strict & Rubric & Strict \\
 & & (n=100) & (n=89) & (n=44) & (n=41) & (n=32) & (n=22) \\
\hline
\multicolumn{8}{l}{\textit{This work (zero-shot, no docs)}} \\
FORM-8B & \textbf{97.7} & \textbf{83.0} & \textbf{18.0} & 43.2 & 14.6 & 56.2 & 50.0 \\
 & [96.5, 98.8] & & [10.1, 25.8] & & [4.9, 26.8] & & [27.3, 68.2] \\
\hline
\multicolumn{8}{l}{\textit{Frontier baselines (with documentation)}} \\
GLM-5.3~\cite{glm5team2026glm53} & 67.8 & 65.0 & 5.6 & \textbf{54.5} & 19.5 & \textbf{65.6} & \textbf{59.1} \\
 & [64.3, 71.4] & & [1.1, 11.2] & & [7.3, 31.7] & & [36.4, 77.3] \\
GLM-5.2 (756B)~\cite{zeng2026glm5} & 75.5 & 55.0 & 5.6 & 31.8 & \textbf{22.0} & 31.2 & 50.0 \\
 & [72.3, 78.6] & & [1.1, 11.2] & & [9.8, 34.1] & & [27.3, 72.7] \\
DeepSeek-V4 Flash (304B)~\cite{xu2026deepseekv4} & 62.3 & 48.0 & 3.4 & 25.0 & 14.6 & 53.1 & 54.5 \\
 & [58.7, 66.1] & & [0.0, 7.9] & & [4.9, 26.8] & & [31.8, 72.7] \\
GPT-OSS-120B (117B)~\cite{agarwal2025gptoss} & 45.6 & 39.0 & 3.4 & 20.5 & 9.8 & 12.5 & 18.2 \\
 & [41.9, 49.4] & & [0.0, 7.9] & & [2.4, 19.5] & & [4.5, 36.4] \\
Gemma 4 (33B)~\cite{deepmind2026gemma4} & 47.6 & 11.0 & 0.0 & 11.4 & 12.2 & 15.6 & 4.5 \\
 & [43.8, 51.4] & & [0.0, 3.3] & & [2.4, 22.0] & & [0.0, 13.6] \\
Qwen3.8-27B~\cite{qwen2026qwen38} & 0.6 & 8.0 & 0.0 & 11.4 & 9.8 & 9.4 & 13.6 \\
 & [0.2, 1.2] & & [0.0, 3.3] & & [2.4, 19.5] & & [0.0, 27.3] \\
Qwen3-8B base~\cite{yang2025qwen3} & 0.0 & 0.0 & 0.0 & 0.0 & 0.0 & 6.2 & 0.0 \\
 & [0.0, 0.5] & & [0.0, 3.3] & & [0.0, 7.1] & & [0.0, 12.7] \\
\hline\hline
\end{tabular}
\end{table}

For zero-count entries the bootstrap percentile interval is degenerate at
zero, so we substitute the one-sided 95\% Clopper--Pearson upper bound
(3.3\%, 7.1\% and 12.7\% for the denominators 89, 41 and 22).

The key findings are:

\begin{enumerate}[leftmargin=*,itemsep=2pt]
  \item On the \textbf{664-Deterministic} benchmark, FORM-8B achieves
  97.7\% [96.5, 98.8], outperforming GLM-5.2 (756B, 75.5\% [72.3, 78.6]) by
  22.2\,pp, GLM-5.3 (67.8\% [64.3, 71.4]) by 29.9\,pp and
  DeepSeek-V4 Flash (304B, 62.3\% [58.7, 66.1]) by 35.4\,pp.
  GPT-OSS-120B (45.6\% [41.9, 49.4]) and Gemma 4 (47.6\% [43.8, 51.4]) score
  even lower, while the Qwen3-8B base model scores 0.0\%---it cannot generate
  a single valid \form{} program without fine-tuning or documentation.
  Qwen3.8-27B is instructive: it produces syntactically valid programs for
  18.8\% of tasks yet matches the expected output on only 0.6\%---the same
  runs-are-not-solutions gap that Section~\ref{sec:true-solve} quantifies on
  the tutorial benchmark.
  The confidence intervals between FORM-8B and all frontier models do not
  overlap, confirming statistical significance.

  \item On the \textbf{Instruct-100} benchmark, FORM-8B leads on both metrics:
  83.0\% [75.0, 90.0] execution rate versus 65.0\% [56.0, 74.0] for the
  strongest frontier model (GLM-5.3), with non-overlapping CIs, and 18.0\%
  [10.1, 25.8] versus 5.6\% [1.1, 11.2] on the strict \form{}-verified
  output-match rate. The strict rate is a conservative lower bound---a
  manual audit found that many Instruct-100 instructions underspecify the
  target computation, so a correct-in-spirit program can still fail the
  check (Section~\ref{sec:true-solve})---but the same grading is applied to
  all models, and the near-disjoint intervals (overlap of 1.1\,pp) make the
  ordering robust. Part of this advantage may reflect our model's
  familiarity with the reference-generation conventions
  (Section~\ref{sec:true-solve}).

  \item On the \textbf{Open-32} benchmark, GLM-5.3 leads on the strict
  metric with 59.1\% [36.4, 77.3] of the 22 output-determined tasks
  versus FORM-8B's 50.0\% [27.3, 68.2] and DeepSeek-V4 Flash's 54.5\%
  [31.8, 72.7]---together with Tutorial-44, the only benchmarks where a
  frontier model outperforms our model on the strict metric, although the CIs
  overlap heavily on this small sample. Section~\ref{sec:true-solve} shows that this lead is
  genuine rather than a rubric artefact: GLM-5.3 converts 13 of its 14
  rubric passes on the output-determined tasks into \form{}-verified
  solutions, while
  our model converts 11 of 11. GLM-5.2 (50.0\%), GPT-OSS-120B (18.2\%) and
  Gemma 4 (4.5\%) score substantially lower. Our v3cn variant
  (Section~\ref{sec:ablation}) ties GLM-5.3 at 59.1\%, and v3c with
  documentation reaches 63.6\%.

  \item On the \textbf{Tutorial-44} benchmark, the strict metric reshuffles
  the ranking: GLM-5.3's apparent dominance under syntax pass (54.5\% vs.\
  our 43.2\%) largely evaporates---it converts only 33\% of its passed
  tasks into genuine solutions, reflecting fluency in writing \emph{runnable}
  \form{} rather than superior problem solving. The strict rates are
  statistically indistinguishable across all strong models (GLM-5.2 22.0\%
  [9.8, 34.1], GLM-5.3 19.5\% [7.3, 31.7], FORM-8B 14.6\% [4.9, 26.8],
  DeepSeek-V4 Flash 14.6\%), and no model solves even a quarter of the
  41 adjudicable tasks: B3-style tutorial transfer remains hard for
  everyone.
\end{enumerate}

\subsection{Runs Are Not Solutions: Strict Output Match}
\label{sec:true-solve}

The execution metric used for B2 and B3 counts a program as passed when it
merely compiles and runs without error---it does not check \emph{what} the
program computes. (B4's rubric judge grades stated criteria, but it is an
LLM assessment rather than an execution check.) The ``Strict'' columns of
Table~\ref{tab:main-results}
grade correctness instead: we re-execute each generated program, extract
its result expressions, and compare them against the reference solution.
Equality is verified by \form{} itself through a difference-to-zero check,
\code{Local Z = (ref) - (gen);}, which is invariant to term ordering,
factor ordering, and algebraic rearrangement (e.g.\ \code{b*b} vs.\
\code{b\^{}2}, or \code{sin(a)*(1-cos(a)\^{}2)} vs.\
\code{sin(a)-sin(a)*cos(a)\^{}2)}). Since instructions do not pin
declaration types, the declaration sets of the reference and the generated
program are tried in turn. A task is \emph{solved} only if every reference
expression is reproduced (extra expressions in the generated output are
tolerated, so printing auxiliary results is harmless). Tasks whose
reference prints no named result expression cannot be adjudicated this way
and are excluded (three on B3, eleven on B2); references reproduce their
own stored outputs on all remaining tasks, validating the metric.

\paragraph*{Instruct-100 instructions are often underspecified.}
Applying strict matching to B2 required a manual audit of the 89
adjudicable tasks, and it exposed a benchmark property worth stating
explicitly: many instructions \emph{underspecify} the target computation,
so the reference output is not uniquely determined by the instruction
alone. In one task the reference multiplies the contraction by a symbol
never mentioned in the instruction; in another, ``apply pattern matching
rules to replace them with simpler expressions'' does not state the rules;
elsewhere ``extract the coefficient of a specific term'' does not say
which term, and ``contract pairs of indices'' admits several valid
readings. A solver that writes a correct, working program for a reasonable
reading of the instruction can therefore still fail the check. The B2
strict column in Table~\ref{tab:main-results} is thus a conservative
\emph{lower bound} on correctness, and the large exec-to-strict gap
(e.g.\ 83.0\% to 18.0\% for our model) mixes genuine errors with
unmatchable specifications. Two remarks qualify our model's B2 strict
lead. The same grading is applied to every model, so the ordering remains
informative, but our training data was generated by
the same pipeline family as the B2 references, so our model may have
learned some of their arbitrary conventions (e.g.\ which symbol to carry
through a contraction); part of its 18.0\% vs.\ 5.6\% lead over the best
frontier model may reflect this familiarity rather than raw skill.

\begin{table}[tp]
\centering
\caption{Execution success vs.\ genuine solutions on Tutorial-44.
``Exec.'' is the syntax-pass rate of Table~\ref{tab:main-results} (n=44);
``Solved'' is the strict \form{}-verified output-match rate on the 41
adjudicable tasks (three references print no result expression). The final
column is the fraction of a model's passed tasks that are genuine solutions.}
\label{tab:true-solve}
\small
\renewcommand{\arraystretch}{1.3}
\begin{tabular}{lccc}
\hline\hline
Model & Exec.\ passed & Solved & Solved/Exec. \\
\hline
GLM-5.2 & 14 (31.8\%) & 9 (22.0\%) & 64\% \\
GLM-5.3 & 24 (54.5\%) & 8 (19.5\%) & 33\% \\
FORM-8B & 19 (43.2\%) & 6 (14.6\%) & 32\% \\
DeepSeek-V4 Flash & 11 (25.0\%) & 6 (14.6\%) & 55\% \\
Gemma 4 & 5 (11.4\%) & 5 (12.2\%) & 100\% \\
Qwen3.8-27B & 5 (11.4\%) & 4 (9.8\%) & 80\% \\
GPT-OSS-120B & 9 (20.5\%) & 4 (9.8\%) & 44\% \\
v3b (ablation, Section~\ref{sec:ablation}) & 10 (22.7\%) & 4 (9.8\%) & 40\% \\
\hline\hline
\end{tabular}
\end{table}

On Tutorial-44, where every task is fully specified by its reference
program, the exec-to-strict conversion is shown in
Table~\ref{tab:true-solve}---and it is sobering for
\emph{every} model: across the board, only a minority of programs that run
also produce the expected result. Two observations stand out. First,
GLM-5.3's apparent lead on B3 largely evaporates: it converts only 33\% of
its passes into genuine solutions, so its strict rate (19.5\%) is
statistically indistinguishable from GLM-5.2's (22.0\%) and from ours
(14.6\%, CIs all overlapping). Its high syntax-pass rate reflects fluency in
writing \emph{runnable} \form{}, not superior problem solving.
Second, models differ markedly in their ``runs-but-wrong'' fraction---from
Gemma 4's 0\% (it passes rarely, but almost always correctly) to 67\% for
GLM-5.3 and 68\% for our own model, the lowest conversion among the strong
models---so syntax-pass alone is a poor proxy for correctness at the
individual-model level. We report both metrics for full transparency,
including for our own model. The strict metric also narrows the absolute
spread between all models to 9.8--22.0\% of the 41 adjudicable tasks,
confirming that B3-style tutorial transfer remains hard for everyone.

\paragraph*{Open-32 under the strict metric.}
Ten of the 32 Open-32 tasks ask the model to define a polynomial ``with at
least four terms'' of its choosing; their expected output is undetermined
and execution-based grading is impossible by construction---this is why B4
uses a rubric judge. On the remaining 22 output-determined tasks we applied
the same \form{}-verified adjudication (the reference programs reproduce
their own outputs on all 22, validating the metric). The rubric judge turns
out to be accurate here: GLM-5.3 converts 13 of its 14 rubric passes on the
output-determined tasks into genuine solutions, DeepSeek-V4 Flash 12 of 13, and our
model 11 of 11. GLM-5.3's Open-32 lead is therefore largely genuine---unlike
its Tutorial-44 lead---although the strict rates (GLM-5.3 59.1\% [36.4,
77.3] vs.\ FORM-8B 50.0\% [27.3, 68.2], $n{=}22$) again have overlapping
CIs, and our ablation variants are on par (v3cn 59.1\%, v3c with
documentation 63.6\%). For GPT-OSS-120B and Gemma 4 the B4 strict column
of Table~\ref{tab:main-results} is based on the no-docs runs, since the
exact generations behind their published rubric scores were not retained.

\paragraph*{Paired significance.}
Exact McNemar tests on the per-task verdicts (Section~\ref{sec:metrics})
confirm what the overlapping CIs already suggest. Our Instruct-100 strict lead
is significant against every frontier model: 16/89 solved versus 5/89 for
GLM-5.3 and GLM-5.2 gives $p = 0.007$, and versus 3/89 for
DeepSeek-V4 Flash and GPT-OSS-120B gives $p < 0.001$. On the small
benchmarks no pairwise comparison among the leading models reaches
significance: on B3, FORM-8B (6/41) vs.\ GLM-5.2 (9/41) gives $p = 0.375$
and GLM-5.2 vs.\ GLM-5.3 (8/41) gives $p = 1.0$; on B4, FORM-8B (11/22)
vs.\ GLM-5.3 (13/22) gives $p = 0.727$ and v3cn vs.\ GLM-5.3 (13/22 each)
gives $p = 1.0$. With 41 and 22 tasks, the entire gap between the best and
worst strong model amounts to one or two discordant tasks, which these
sample sizes cannot resolve: B3 and B4 can establish that \emph{every}
model is far from solving them, but not that any model leads.

\subsection{Ablation Studies}
\label{sec:ablation}

\begin{figure}[t]
  \centering
  \includegraphics[width=0.85\textwidth]{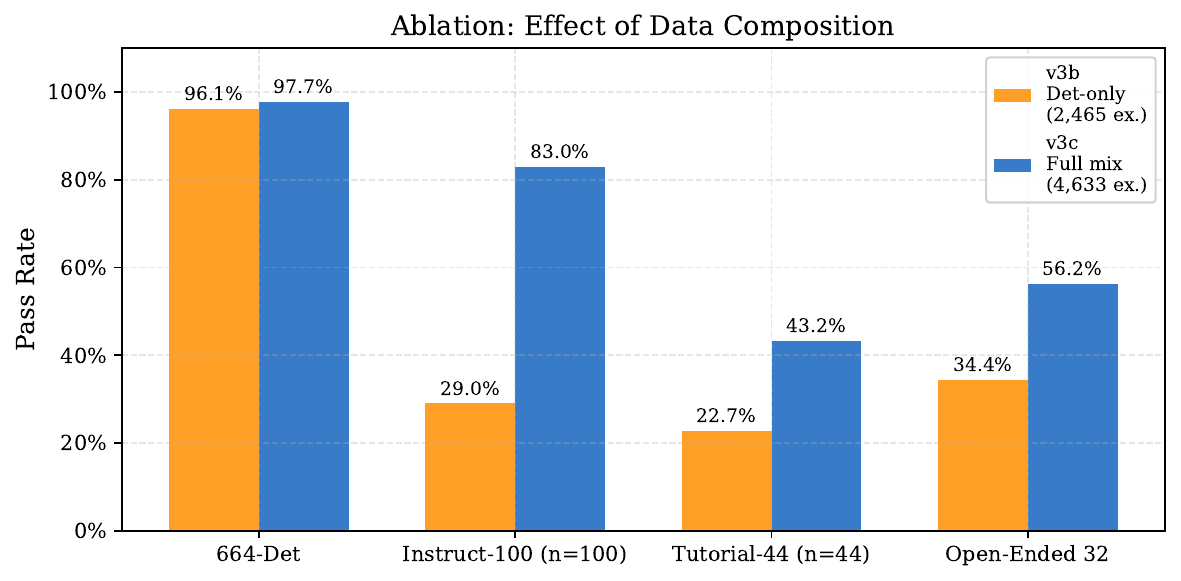}
  \caption{Ablation: effect of data composition. The v3b model
  (deterministic-only data, 2{,}465 examples) collapses open-ended capability.
  The v3c model (full mix, 4{,}633 examples) recovers it while maintaining
  deterministic performance.}
  \label{fig:ablation}
\end{figure}

\paragraph*{Data Composition.}
Figure~\ref{fig:ablation} shows the effect of data composition across
benchmarks. The v3b model (2{,}465 verified deterministic examples) achieves
excellent deterministic performance (96.1\%) but \emph{collapses} on
open-ended tasks (29.0\% on Instruct-100), demonstrating that deterministic-only
training narrowly specialises the model and destroys its ability to generate
code from open-ended instructions. Adding audited open-ended code, tutorial
programs, and knowledge Q\&A (v3c, 4{,}633 examples) recovers open-ended
capability (83.0\% on Instruct-100) while \emph{improving} deterministic
performance (97.7\%). A further variant, v3cn, trained on the same
4{,}633 examples with the documentation-in-context prefixes removed, is
marginally stronger on B2 (88.0\%) and B4
(68.8\%) but slightly weaker on B1 (97.0\%) and B3 (29.5\%); we retain the
documentation slice in the released model because it matches deployment
conditions in which users may paste reference material into the prompt.

\begin{figure}[t]
  \centering
  \includegraphics[width=0.8\textwidth]{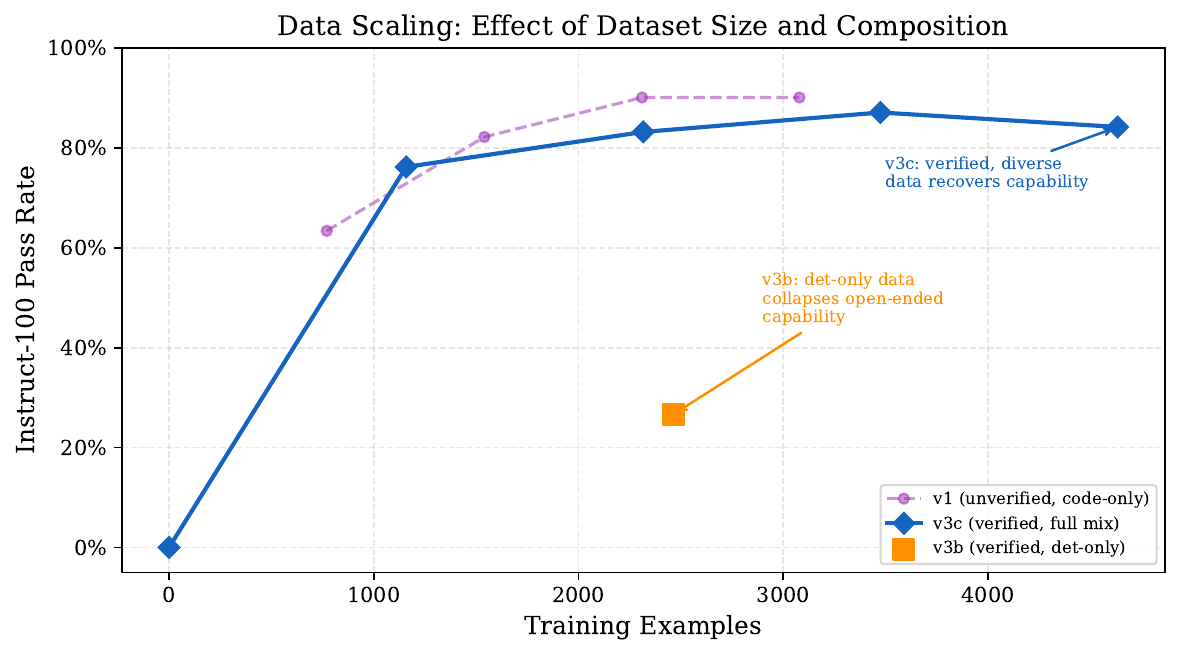}
  \caption{Data scaling: effect of dataset size and composition. The v1 scaling
  curve (purple, dashed) shows improvement with more unverified code-only data.
  The v3c curve (blue, solid) shows the effect of verified, diverse data at
  25\%/50\%/75\%/100\% of the full dataset. The v3b point (orange square)
  shows that verified but deterministic-only data \emph{collapses} open-ended
  capability despite having more examples than the 50\% v3c subset.}
  \label{fig:scaling}
\end{figure}

\paragraph*{Data Scaling.}
Figure~\ref{fig:scaling} shows the effect of dataset size and composition on
Instruct-100 pass rate (all numbers in this paragraph were measured on an
earlier snapshot of the benchmark; model rankings are unchanged). We compare two data-generation strategies: (1)~the v1
pipeline (the first, unverified iteration of our data pipeline: code-only
data), which scales from 770 to 3{,}079
examples and reaches 90.1\% at 75\% of the data, and (2)~the v3c pipeline
(verified, diverse data), which scales from 0 to 4{,}633 examples. The v3c
ablation reveals a steep learning curve: the base model scores 0.0\%, but
with just 1{,}158 verified examples (25\%), pass rate jumps to 76.2\%, and
with 2{,}316 examples (50\%) it reaches 83.2\%---nearly matching the full
model's 84.2\% on the same benchmark. The v3b model (2{,}465 \emph{verified} deterministic-only
examples) scores only 26.7\%, demonstrating that data \emph{composition}
matters more than raw quantity: 2{,}316 diverse verified examples (83.2\%)
dramatically outperform 2{,}465 deterministic-only verified examples (26.7\%).
This confirms that verified multi-source data is the key driver of
performance.

\paragraph*{Documentation in Context.}
Because frontier baselines are evaluated with a syntax guide in context
(Section~\ref{sec:baselines}), one might suspect that our main-table advantage
is a documentation artefact. It is not: evaluated \emph{with} the same guide
that the frontier models receive, our top models barely move (v3c
83.0$\to$88.0\%, v3cn 88.0$\to$85.0\% on the 100-task benchmark), whereas
weak models are rescued substantially (v3b 29.0$\to$62.0\%; the
un-fine-tuned base 0.0$\to$34.0\%). Fine-tuning subsumes the information
carried by the documentation; documentation alone cannot substitute for it
at the level of our best models, though it explains part of the spread among
weaker baselines.

\subsection{Capability Preservation}
\label{sec:capability}

\begin{figure}[t]
  \centering
  \includegraphics[width=0.68\textwidth]{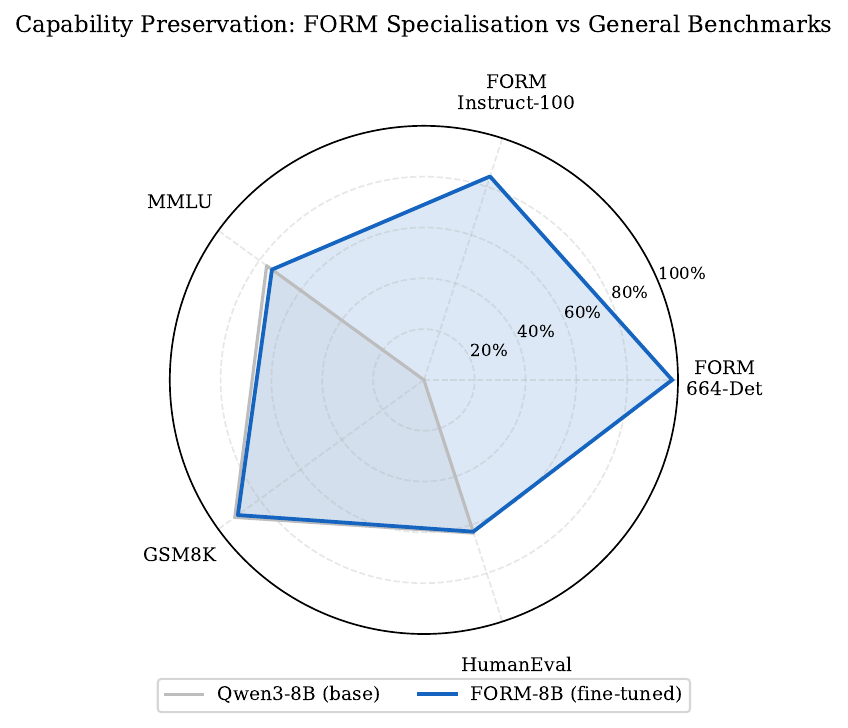}
  \caption{Capability radar: the fine-tuned model (blue) gains dramatically
  on \form{} benchmarks while retaining general capabilities (MMLU, GSM8K,
  HumanEval) close to the base model (grey).}
  \label{fig:radar}
\end{figure}

A key concern with fine-tuning is \emph{catastrophic forgetting} of
pre-trained capabilities. We verify that QLoRA fine-tuning preserves general
reasoning by evaluating on three standard benchmarks
(Table~\ref{tab:capability} and Figure~\ref{fig:radar}):

\begin{table}[t]
\centering
\caption{General capability preservation. Changes relative to Qwen3-8B base.
MMLU: full 9{,}183-question test set; GSM8K: 200-problem subsample;
HumanEval: all 164 problems.}
\label{tab:capability}
\renewcommand{\arraystretch}{1.3}
\begin{tabular}{lccc}
\hline\hline
Benchmark & Base & FORM-8B & $\Delta$ \\
\hline
MMLU (5-shot)~\cite{hendrycks2021mmlu} & 76.5\% & 73.9\% & $-2.6$\,pp \\
GSM8K (0-shot, flexible-extract)~\cite{cobbe2021gsm8k} & 92.0\% & 90.5\% & $-1.5$\,pp \\
HumanEval (0-shot)~\cite{chen2021codex} & 63.4\% & 62.8\% & $-0.6$\,pp \\
\hline\hline
\end{tabular}
\end{table}

The minimal degradation ($\leq 2.6$\,pp on all three benchmarks) confirms
that QLoRA---which trains only 0.5\% of the total parameters---effectively
prevents catastrophic forgetting. This is consistent with the
literature~\cite{dettmers2023qlora, hu2022lora}.

\subsection{Agentic Workflow Modification}
\label{sec:agentic}

The benchmarks in the previous sections evaluate single-turn code
generation: one instruction, one program, one verdict. In practice, however,
physicists use \form{} as part of multi-file computational workflows that
include diagram generation (e.g., \textsc{Qgraf}~\cite{nogueira2003qgraf}),
shell orchestration scripts, and hand-written \form{} procedures. A
practically useful AI assistant must therefore operate in an \emph{agentic}
mode: given a repository of undocumented code, it must read files, understand
the calculation pipeline, and rewrite it to accomplish a modified physics
goal.

Robust comparison of different LLMs in agentic mode is difficult for several
reasons. First, each model benefits from a different system prompt and
tool-use style; a prompt optimised for one model may disadvantage another.
Second, agentic runs are inherently stochastic---the same model may take
different code paths across runs---so many repetitions are needed for
reliable statistics. Third, the evaluation metric (``did the rewritten
workflow produce the correct physics?'') requires running the full pipeline
and comparing output files, making it expensive and task-specific.
These obstacles make a fair, large-scale agentic benchmark beyond the scope
of this work.

Nevertheless, to illustrate the practical value of our fine-tuned model, we
report a representative user experience using the \textsc{Pi} agent
harness~\cite{zechner2025pi}---a minimal, extensible terminal-based coding
agent. In this demonstration, an agent is given an undocumented \form{}
codebase that computes the tree-level squared matrix element for $gg \to
t\bar{t}$, along with the \textsc{Qgraf} configuration for amplitude generation
and an orchestrating bash script. The agent is tasked with extending the
workflow to compute the virtual NLO correction---specifically, the interference
between the Born and one-loop amplitudes,
$2\,\mathrm{Re}(\mathcal{M}^{(0)*}\mathcal{M}^{(1)})$.

We compare two configurations of the agent:
\begin{itemize}[leftmargin=*,itemsep=2pt]
  \item \textbf{Smart-routed:} The agent uses Qwen3.8-27B as the primary
  model, with a routing rule that directs any request involving \form{} code
  to our fine-tuned 8B model. Non-\form{} tasks (editing bash scripts,
  modifying \textsc{Qgraf} configurations, writing documentation) are handled by
  the 27B model. No \form{} documentation is included in the context.
  \item \textbf{Frontier with docs:} The agent uses a single frontier model
  (GLM-5.2, 756B parameters) for all tasks, with the first 6{,}000 characters
  of the \form{} syntax guide prepended to every prompt.
\end{itemize}

Both configurations are given the same repository and the same natural-language
goal, with no additional skills or extensions enabled in the harness. The
smart-routed configuration successfully completes the task in fewer agent
steps and with substantially fewer total input/output tokens, because the
fine-tuned model generates correct \form{} code without needing to consult
documentation on every turn. The frontier-with-docs configuration, by
contrast, requires multiple repair attempts on the \form{} portions of the
workflow, increasing both token consumption and wall-clock time. While this
is a single demonstration rather than a controlled study, it illustrates the
practical advantage of domain-specialised models in agentic physics
calculations: routing \form{}-related subtasks to a small specialist avoids
the overhead of injecting documentation into every frontier-model prompt,
while achieving higher code quality.


\section{Discussion}
\label{sec:discussion}

\subsection{Why Does an 8B Model Outperform 756B?}
\label{sec:why}

The result that an 8B fine-tuned model outperforms a 756B model by
$>$20\,pp may seem surprising. The explanation lies in the
\emph{zero-resource nature} of \form{}: the 756B model has never seen
\form{} code during pre-training, and even with 6{,}000 characters of syntax
documentation, it must \emph{infer} \form{}'s syntax from a brief description.
The fine-tuned 8B model, by contrast, has been trained on 4{,}633 verified
examples and has internalised \form{}'s syntax, idioms, and common patterns.
This demonstrates a key insight for scientific domain-specific languages:
\emph{verified task-specific data trumps raw parameter count} when the target
language is absent from pre-training corpora.

\subsection{Absence of Proprietary API Models}
\label{sec:no-api}

A natural question is why we do not include results for proprietary API
models such as ChatGPT (GPT-4/5), Claude, or Gemini. Our evaluation
pipeline---which executes generated programs through the \form{} binary and
compares outputs---is fully compatible with any model that accepts a text
prompt and returns a text completion. However, evaluating these models
requires paid API access, and the scale of our benchmark suite (664 + 100 +
44 + 32 = 840 single-attempt evaluations per model) would incur significant
cost. Moreover, proprietary models frequently update their weights without
version pinning, making results non-reproducible. We therefore restrict our
baseline comparisons to open-weights models that can be self-hosted,
version-pinned, and evaluated with identical hardware and software
configurations, ensuring full reproducibility.

\subsection{Limitations}
\label{sec:limitations}

We identify several limitations of the current work:

\begin{enumerate}[leftmargin=*,itemsep=2pt]
  \item \textbf{Loosely-specified and small open-ended test sets}: The strict
  output-match metric is well-defined only for a subset of the open-ended
  benchmarks (89 of 100 B2 tasks; 22 of 32 B4 tasks), and a manual audit
  found that many B2 instructions underspecify the target computation, so
  the B2 strict rate is a lower bound. The adjudicable subsets are also
  small (41 tasks for B3, 22 for B4): confidence interval half-widths reach
  $\pm$20\,pp,
  so results there are suggestive rather than significant. Expanding the
  open-ended test sets to 100+ fully-specified tasks would strengthen these
  claims.

  \item \textbf{No human evaluation}: We do not conduct a user study with
  physicists. Execution-based metrics provide objective signal, but a human
  evaluation would validate the practical utility of the model.

  \item \textbf{Single base model}: We fine-tune only Qwen3-8B. Results may
  differ for other base models (e.g., Qwen3-Coder, Llama).

  \item \textbf{Author-constructed benchmarks}: The evaluation benchmarks and
  the fine-tuned model were developed by the same author, which inevitably
  introduces a potential bias: the benchmark tasks may inadvertently reflect
  the author's own programming style and expectations, and may align with the
  training data used to fine-tune the model. In the absence of a
  community-established benchmark for \form{} code generation---a gap that
  reflects the zero-resource status of the language---constructing an in-house
  benchmark was the only viable option. We try to mitigate this concern by
  grounding all benchmark tasks in the official \form{} documentation and real
  physics use cases and by relying on execution-based metrics against the
  \form{} binary rather than subjective assessment. Releasing the suite as a
  public benchmark (Section~\ref{sec:conclusion}) allows the community to
  audit and extend it.

  \item \textbf{Correctness, not runtime performance}: All metrics verify
  whether generated code produces the correct output, not how fast it
  runs. Optimising \form{} runtime (statement ordering, \code{.sort}
  placement, dollar variables, pattern constraints) is a distinct expert
  skill in its own right; the training corpus carries no runtime-optimality
  signal and the benchmarks measure none. A runtime-aware extension (e.g.,
  rewarding programs that beat a reference implementation's wall-clock
  time on large expressions) is a natural follow-up.
\end{enumerate}

\subsection{Future Work}
\label{sec:future}

Several directions are promising for future work:

\begin{enumerate}[leftmargin=*,itemsep=2pt]
  \item \textbf{Reinforcement Learning from Verifiable Rewards (RLVR)}: The
  \form{} binary provides a deterministic reward signal (does the program run
  and produce the expected output?). This makes it a natural fit for GRPO
  (Group Relative Policy Optimization~\cite{shao2024grpo}), which has shown
  +9--16\,pp improvements for code models
  (cf.~\cite{guo2025deepseekr1}).

  \item \textbf{Expanded open-ended test set}: Increasing the Open-32 and
  Tutorial-44 benchmarks to 100+ tasks would reduce confidence interval widths
  and enable more definitive comparisons on open-ended generation.

  \item \textbf{Human evaluation}: A study with 2--3 physicists evaluating
  50 generated programs would validate practical utility.

  \item \textbf{Generalisation to other scientific DSLs}: The
  verification-driven pipeline is applicable to any language with a
  deterministic execution environment (e.g.,
  Cadabra~\cite{peeters2007cadabra},
  Redberry~\cite{bolotin2013redberry}, and
  REDUCE~\cite{hearn1973reduce}).
\end{enumerate}


\section{Conclusion}
\label{sec:conclusion}

We have demonstrated that a small (8B) language model, fine-tuned with QLoRA
on 4,633 verification-driven examples, can generate semantically correct
\form{} symbolic algebra code at a level that significantly exceeds
frontier-scale models (up to 756B parameters) on the benchmarks large enough
to decide the comparison. The key insight is that for
zero-resource domain-specific languages, verified task-specific training data
is more valuable than raw parameter count. Our verification-driven
pipeline---which uses the \form{} binary as an execution oracle---ensures that
every training example is syntactically valid and semantically consistent,
producing high-quality data without human annotation.

The resulting model achieves 97.7\% exact output match on 664 deterministic
tasks and an 83.0\% execution rate on 100 open-ended tasks, with
non-overlapping 95\% confidence intervals against GLM-5.3, GLM-5.2 (756B)
and DeepSeek-V4 Flash (304B) on these benchmarks. The advantage persists
under strict \form{}-verified output matching: on the output-determined
Instruct-100 subset (the 89 tasks with uniquely determined outputs) our
model solves 18.0\% versus 5.6\% for the best frontier
model, and exact paired McNemar tests confirm the gap against every
frontier model ($p \le 0.007$). On the small Tutorial-44 and Open-32
benchmarks, where no model solves even a quarter of the tasks, differences
of the size observed between the leading models are statistically
unresolvable ($p \ge 0.375$): these benchmarks show that \emph{everyone}
struggles, but not that any model leads. General reasoning capability is
preserved to within 2.6\,pp on MMLU, GSM8K, and HumanEval, confirming that
QLoRA fine-tuning does not cause catastrophic forgetting.

This work provides a template for bringing AI-assisted coding to niche
scientific languages: identify a deterministic execution oracle, generate
and verify training data through it, and fine-tune a small model with
parameter-efficient methods. As \form{} continues to underpin precision
calculations in particle physics~\cite{davies2026form5}, AI tooling that
lowers the barrier to its use becomes increasingly important for the
community.

The fine-tuned model weights and the complete benchmark suite---840 tasks
with instructions, references, evaluation harness, and per-task verdicts for
all evaluated models---are publicly available on HuggingFace:
the \textsc{FORM-8B} model~\cite{chargeishvili2026form8b} and the
\textsc{FORM-bench} benchmark suite~\cite{chargeishvili2026formbench},
each with a persistent Digital Object Identifier. The training corpus is
available from the author upon request.

\subsection*{Acknowledgements}
The author thanks Gudrun Heinrich, Adam Kardos, Marco Klann, Vitaly
Magerya and Sven-Olaf Moch for discussions and comments on the manuscript.

The work of B.C. was supported by the Deutsche Forschungsgemeinschaft (DFG)
under grant 396021762 --- TRR 257 {\it ``Particle Physics Phenomenology after
the Higgs Discovery''}.

\appendix


\section{Representative Inference Examples}
\label{sec:appendix-examples}

The following examples are drawn from the Instruct-100 benchmark and were
generated by the v3c model (FORM-8B) in a single attempt at temperature\,0.0.
Each example shows the natural-language instruction, the generated \form{}
program, and the actual \form{} binary output. All programs were verified by
running through FORM\,5.0.

\begin{figure}[ht]
\begin{verbatim}
Indices mu, nu;
Vectors p, q;
Dimension 4;
Local E = g_(1, mu) * g_(1, nu) * p(mu) * q(nu);
Trace4, 1;
contract;
Print;
.end
\end{verbatim}
\caption{\textbf{Example A1: Dirac trace with momentum contraction.}
Instruction: ``Compute the trace of a product of gamma matrices contracted
with two external vectors, and simplify the result using the metric tensor in
four dimensions.'' The model correctly uses \code{Trace4} for the Dirac trace,
\code{contract} for Lorentz index contraction, and sets \code{Dimension 4}.
FORM output: \texttt{E = 4*p.q}, reflecting the trace identity
$\mathrm{Tr}[\gamma^\mu \gamma^\nu] = 4\,g^{\mu\nu}$.}
\label{fig:ex-a1}
\end{figure}

\begin{figure}[ht]
\begin{verbatim}
Indices mu, nu, rho, sig;
Set ind: mu, nu, rho, sig;
CTensor g;
Local E = g(mu, nu) * g(nu, rho)
           * g(rho, sig) * g(sig, mu);
id g(mu?, nu?) = d_(mu, nu);
contract;
Print;
.end
\end{verbatim}
\caption{\textbf{Example A2: Wildcard pattern matching for metric contraction.}
Instruction: ``Contract pairs of indices from a declared index set using the
metric tensor, then print the resulting simplified scalar expression.'' The
model uses the wildcard pattern \code{g(mu?, nu?)} to replace the symbolic
tensor \code{g} with the metric \code{d\_} and then contracts. FORM output:
\texttt{E = 4}, the correct result for a chain of four metric contractions in
four dimensions.}
\label{fig:ex-a2}
\end{figure}

\begin{figure}[ht]
\begin{verbatim}
Symbols a, b, c;
Functions f (symmetric), g (antisymmetric);
Local E = f(a,b) + g(a,b) + f(b,a) + g(b,a);
id f(a,b) = a + b;
id g(a,b) = a - b;
Print;
.end
\end{verbatim}
\caption{\textbf{Example A3: Symmetric and antisymmetric function properties.}
Instruction: ``Declare a symmetric function and an antisymmetric function,
then apply pattern matching rules to replace them with simpler expressions and
print the result.'' The model correctly declares the symmetry properties,
so that \code{f(b,a)} is automatically identified with \code{f(a,b)} and
\code{g(b,a)} with \code{-g(a,b)}. FORM output: \texttt{E = 2*b + 2*a},
corresponding to $2(a+b) + (a-b) + (b-a) = 2a+2b$.}
\label{fig:ex-a3}
\end{figure}

\begin{figure}[ht]
\begin{verbatim}
Symbols x;
Local E = 0;
#do i = 1, 5
    Local E = E + x^`i';
#enddo
Print +f;
.end
\end{verbatim}
\caption{\textbf{Example A4: Preprocessor loop and the module-lifecycle
pitfall.}
Instruction: ``Use a preprocessor loop to build a sum of powers of a symbol,
then print the resulting polynomial.'' The model correctly uses the
\code{\#do}/\code{\#enddo} preprocessor loop with backtick variable
substitution, and the program runs without error. The resulting output is
nevertheless \emph{not} the requested sum: because the \code{Local}
statement redefines \code{E} within a single module (there is no
\code{.sort} between loop iterations), each iteration discards the previous
partial sum, and the final output is $E = x^5$ rather than
$x + x^2 + x^3 + x^4 + x^5$. We deliberately present this as an instructive
failure example: the missing \code{.sort} is exactly the subtle \form{}
idiom discussed in Section~\ref{sec:background}.}
\label{fig:ex-a4}
\end{figure}

\begin{figure}[ht]
\begin{verbatim}
Symbols p2, m2, A, B, C;
Vectors p, q;
Indices mu, nu;
CTensor g;
Local F = g(mu,nu)*A + p(mu)*p(nu)*B
          + q(mu)*q(nu)*C;
id g(mu,nu) = d_(mu,nu);
id p(mu)*p(nu) = p2*g(mu,nu);
id q(mu)*q(nu) = m2*g(mu,nu);
contract;
Print +f;
.end
\end{verbatim}
\caption{\textbf{Example A5: Tensor decomposition with metric substitution.}
Instruction: ``Decompose a rank-2 tensor integral with two Lorentz indices into
scalar integrals by contracting with the metric tensor and momentum vectors,
then simplifying the result.'' The model constructs a general rank-2 tensor
decomposition, then applies \code{id} rules to replace tensor structures with
scalar coefficients. FORM output:
\texttt{d\_(mu,nu)*A + g(mu,nu)*m2*C + g(mu,nu)*p2*B}.}
\label{fig:ex-a5}
\end{figure}

\clearpage

\bibliographystyle{JHEP}
\bibliography{references}

\end{document}